\documentclass[twoside,final]{pro12}

\usepackage{amsmath}

\usepackage[pdftex]{graphicx} 

\head{J.\,D. Turner et al.} {LOFAR low-frequency beam-formed observations of exoplanets}
\begin{document}

\title{The search for radio emission from exoplanets using LOFAR low-frequency beam-formed observations: Data pipeline and preliminary results for the 55 Cnc system}
\author{J.\,D. Turner\adress{\textsl Department of Astronomy, University of Virginia, Charlottesville, VA, USA}$\,$ \adress{\textsl Laboratoire de Physique et Chemie de l'Environment et de l'Espace (LPC2E) Universit\'{e} d'Orl\'{e}ans/CNRS, Orl\'{e}ans, France}$\,$, J.--M. Grie{\ss}meier$^\dagger$\adress{\textsl Nan\c{c}ay Radio Observatory, Observatoire de Paris-CNRS/INSU, Meudon, France}$\,$, P. Zarka\adress{\textsl LESIA \& USN, Observatoire de Paris/CNRS/PSL, Meudon, France}$\,$, and I. Vasylieva\adress{\textsl Institute of Radio Astronomy, National Academy of Sciences of Ukraine, Kharkov, Ukraine}}

\maketitle

\begin{abstract}
Detection of radio emission from exoplanets can provide information on the star-planet system that is difficult to study otherwise, such as the planetary magnetic field, magnetosphere, rotation period, interior structure, atmospheric dynamics and escape, and any star-planet interactions. Such a detection in the radio domain would open up a whole new field in the study of exoplanets. However, currently there are no confirmed detections of an exoplanet at radio frequencies. In this study, we search for non-thermal radio emission from the 55 Cnc system which has 5 known exoplanets. According to theoretical predictions 55~Cnc~e, the innermost planet, is among the best targets for this search. We observed for 18 hours with the Low-Frequency Array (LOFAR) Low Band Antenna in the frequency range 26-73 MHz with full-polarization and covered 85$\%$ of the orbital phase of 55~Cnc~e. During the observations four digital beams within the station beam were recorded simultaneously on 55 Cnc, nearby ``empty'' sky, a bright radio source, and a pulsar. A pipeline was created to automatically find and mask radio frequency interference, calibrate the time-frequency response of the telescope, and to search for bursty planetary radio signals in our data. Extensive tests and verifications were carried out on the pipeline. Analysis of the first 4 hours of these observations do not contain any exoplanet signal from 55 Cnc but we can confirm that our setup is adequate to detect faint astrophysical signals. We find a 3-sigma upper limit for 55 Cnc of 230 mJy using the pulsar to estimate the sensitivity of the observations and 2.6 Jy using the time-series difference between the target and sky beam. The full data set is still under-going analysis. In the near future we will apply our observational technique and pipeline to the most promising exoplanet candidates for which LOFAR observations have already been obtained. 

\end{abstract}

\section{Introduction}

One of the most elusive goals in exoplanet science today is the detection of exoplanetary magnetic fields. Observations of an exoplanet magnetosphere would allow constraints on planetary properties difficult to study such as their magnetic field strength and structure, rotation period, interior structure, atmospheric dynamics and escape, the presence of extrasolar moons, and the physics of star-planet interactions (Hess $\&$ Zarka 2011). The question of whether intrinsic fields, like those at Jupiter and Saturn, are present on gas giant exoplanets is critical because it greatly affects our understanding of their origins and evolution. Additionally, the deflection of stellar wind particles and cosmic rays due to Earth's magnetic field help contribute to its habitability and this may also be the case for exoplanets (e.g. Grie{\ss}meier 2015). 

The most promising method to detect exoplanet magnetic fields is cyclotron radio emission observations because this method is not susceptible to false positives (Grie{\ss}meier 2015, and references therein). However, many studies conducted to find exoplanet radio emission have resulted in non-detections (Zarka et al. 2015, Grie{\ss}meier 2015, Grie{\ss}meier 2017, and references therein). A few studies find potential detections (Lecavelier~des~Etangs et al. 2013, Sirothia et al. 2014) but they remain unconfirmed. In this study, we will use for the first time the Low-Frequency Array (LOFAR) Low Band Antenna (LBA) in beam-formed mode to search for radio emission from exoplanets. 

\subsection{Predictions for radio emission from 55~Cnc~e} \label{sec:predict}
A large amount of theoretical work has been done on predicting radio emission fluxes and maximum frequencies for exoplanets (Zarka et al. 2015, Grie{\ss}meier 2015, Grie{\ss}meier 2017, and references therein). 55 Cnc was determined to be one of the best targets for radio observations due to advantages of a small orbital distance for 55~Cnc~e (the inner-most planet), proximity, and planetary multiplicity (Grie{\ss}meier et al. 2007) and it shows hints of radio variability in UTR-2 data (V. Ryabov, personal communication). Theoretical predictions suggest the existence of decameter emission up to a few tens of MHz for 55~Cnc~e and corresponding flux densities up to hundreds of mJy (Grie{\ss}meier et al., 2007, Nichols et al. 2016). Additionally, 55~Cnc~e is a transiting planet which will allow for the possibility of observing a planetary occultation in the radio domain (as done in Lecavelier~des~Etangs et al. 2013).

\section{LOFAR Observations} \label{sec:LOFAR}
We observed for 18 hours with LOFAR LBA (van~Haarlem et al. 2013) in the frequency range 26-73 MHz with full-polarization in beam-formed mode. The observational setup can be found in Table \ref{tb:setup}. The observations were performed during night/dawn time hours in order to avoid strong contamination by radio frequency interference (RFI). During the observations four digital core beams (FWHM: 7 arcmins at 60 MHz) within the station beam (FWHM: 10 degrees at 60 MHz) were recorded simultaneously on (1) 55 Cnc, (2) a patch of nearby “empty” sky, (3) the nearby pulsar B0823+26, and (4) a bright radio source (0858.1+2750; 30 Jy at 60 MHz). The extra beams make this setup unique since they can be used for control of instrumental effects, verify that a detection in the exoplanet beam is not a false positive detection (e.g. ionospheric fluctuations), and check the reliability of the data-reduction pipeline. Additionally, cyclotron radio emission observations are expected to be strongly circularly polarized and therefore the polarization information can be also used to verify a real signal. The theoretical sensitivity of the LOFAR observations is $\sim$16 mJy using the entire bandwidth and a 2-minute integration\footnote{The sensitivity ($\Delta S$) was calculated using the sensitivity equation $\Delta S= S_{sys}\alpha/\sqrt{N(N-1)n_{pol}b \tau}$ where $S_{sys}$ is the system equivalent flux density (SEFD) of an LBA core station (40 kJy, obtained from LOFAR calibration data; van~Haarlem et al. 2013), $N$ is the number of stations used, $n_{pol}$ is the number of polarizations (2), $b$ is the bandwidth, $\tau$ is the total time of observation, and $\alpha$ is a factor (equal to 1 for the calculated value) taking into account incoherent addition and flagging of data (see Section \ref{sec:discussion}).}.  

\begin{table}[!htb]
\centering
\caption{Setup of LOFAR observations of 55 Cnc}
\begin{tabular}{cc}
\hline 
\hline
 Starting Frequency (MHz)        & 26     \\
 Ending Frequency (MHz)          & 73     \\
 Frequency Resolution (kHz)     & 3.05      \\
 Number of Subbands                       & 244       \\
 Channels per Subband           &64             \\
 Time Resolution (msec)         & 10.5      \\
 Total Observing Time (hours)             & 4.38 \\
Number of LOFAR stations        & 24 (core)      \\
 Beams                          & target, pulsar, sky, bright source\\
 Polarizations                  & IQUV   \\
\hline
\end{tabular}
\label{tb:setup}
\end{table}

\begin{figure}[ht!]
 \center
 \begin{tabular}{cc}
 {\put(0,90){(A)}\includegraphics[width=0.50\textwidth,page=1]{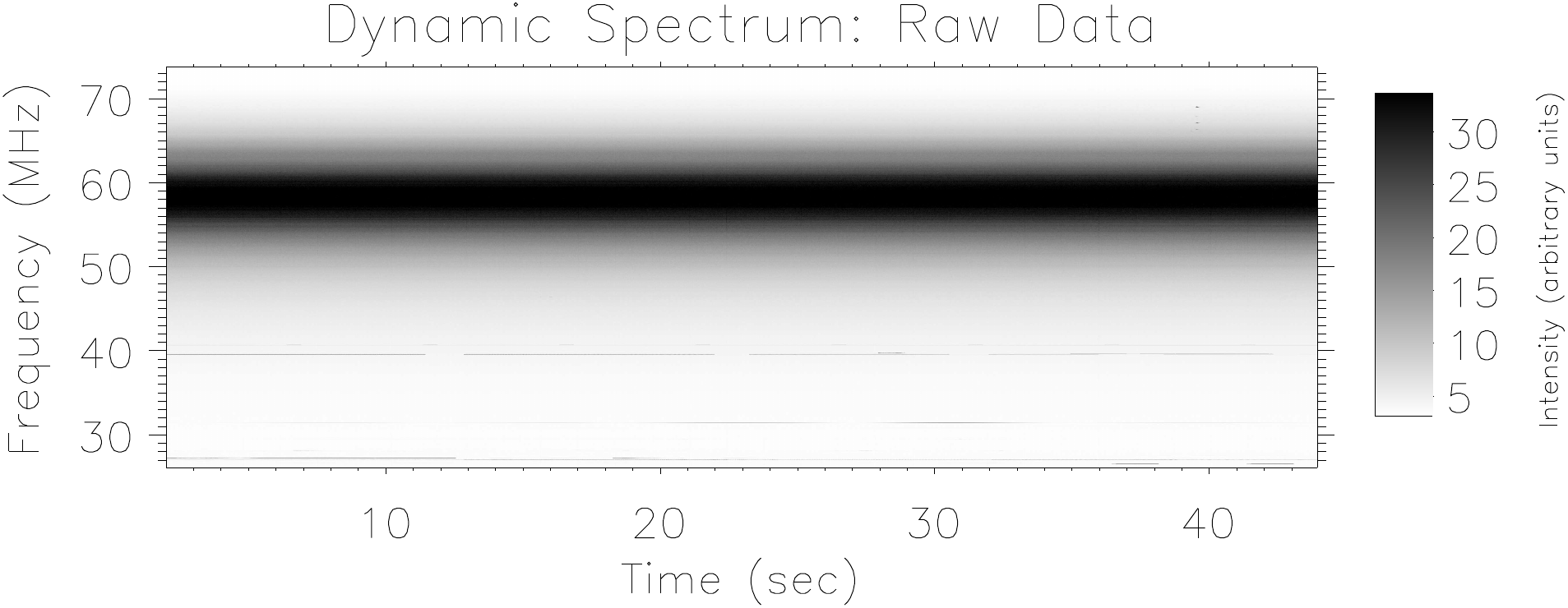}} &
{\put(0,90){(B)} \includegraphics[width=0.50\textwidth,page=3]{Figure1.pdf}}\\
{\put(0,90){(C)}\includegraphics[width=0.50\textwidth,page=2]{Figure1.pdf}}&
{\put(0,90){(D)}\includegraphics[width=0.50\textwidth,page=4]{Figure1.pdf}}\\
 \end{tabular}
  \caption{\textbf{(A)} Example dynamic spectrum of raw Obs~$\#$1 LOFAR data. \textbf{(C)} Zoomed in dynamic spectrum of panel A. Time series (\textbf{B}) and integrated spectrum (\textbf{D}) of the raw Obs~$\#$1 LOFAR data. The pronounced peak of the frequency response function at 58 MHz is easily seen in the dynamic spectrum and the integrated spectrum. RFI is also easily identifiable as bright spikes in all plots.}
  \label{fig:raw}
\end{figure}

\section{Data Pipeline for LOFAR Observations} 
\begin{figure}[ht]
 \begin{center}
\includegraphics[width=0.90\textwidth]{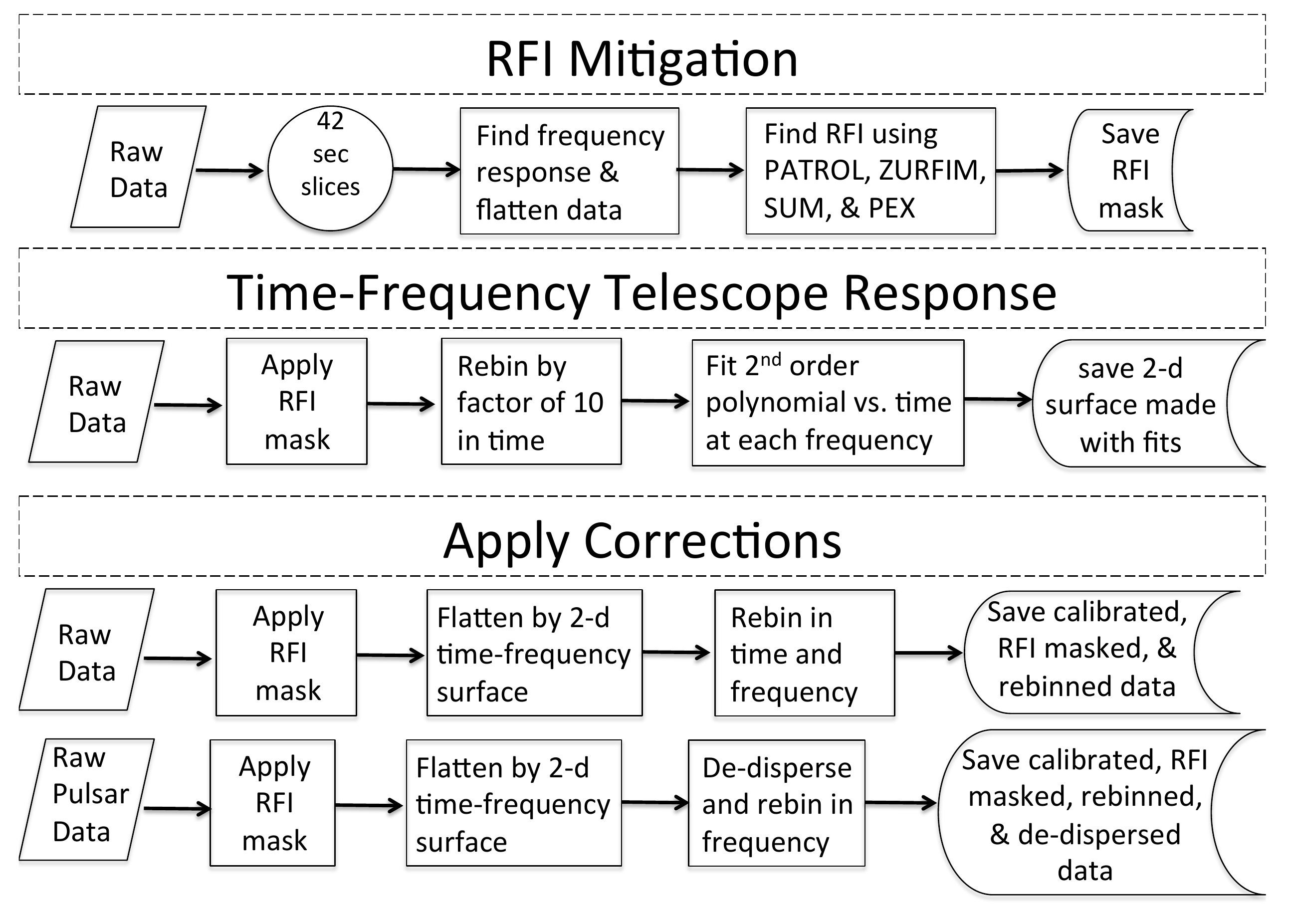}
\caption{Flow chart of the main parts of the pipeline: RFI mitigation, finding the time-frequency telescope response, and applying these corrections to the data.}
\label{fig:flowchart}
\end{center}
\end{figure}

We created a pipeline that automatically calibrates the data for instrumental effects and finds and masks RFI. This pipeline was adapted from the one created by Vasylieva (2015) to search for radio emission from exoplanets using UTR-2. A flow chart of the pipeline can be found in Figure \ref{fig:flowchart}. The pipeline consists of three main parts: RFI mitigation, finding the time-frequency telescope response, and applying the corrections found in the first two steps to the data. We will describe each part of the pipeline in greater detail below.

\subsection{RFI Mitigation} \label{sec:RFI_mask}
\begin{figure}[htb!]
 \begin{center}
 \begin{tabular}{cc}
 \vspace{-0.2em}
 {\put(0,170){(A)}\includegraphics[width=0.5\textwidth]{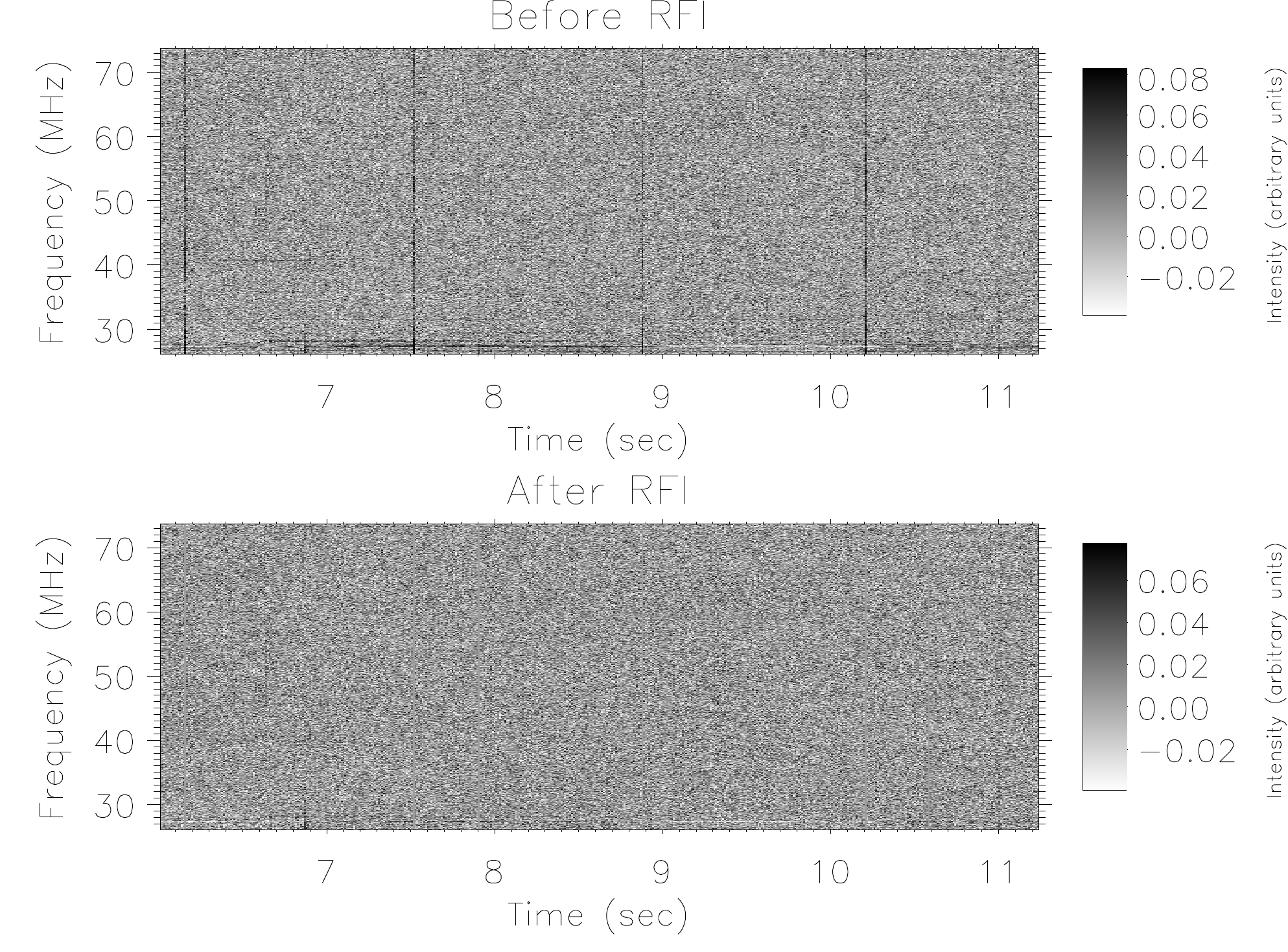}} & 
{\put(0,170){(B)} \includegraphics[width=0.5\textwidth]{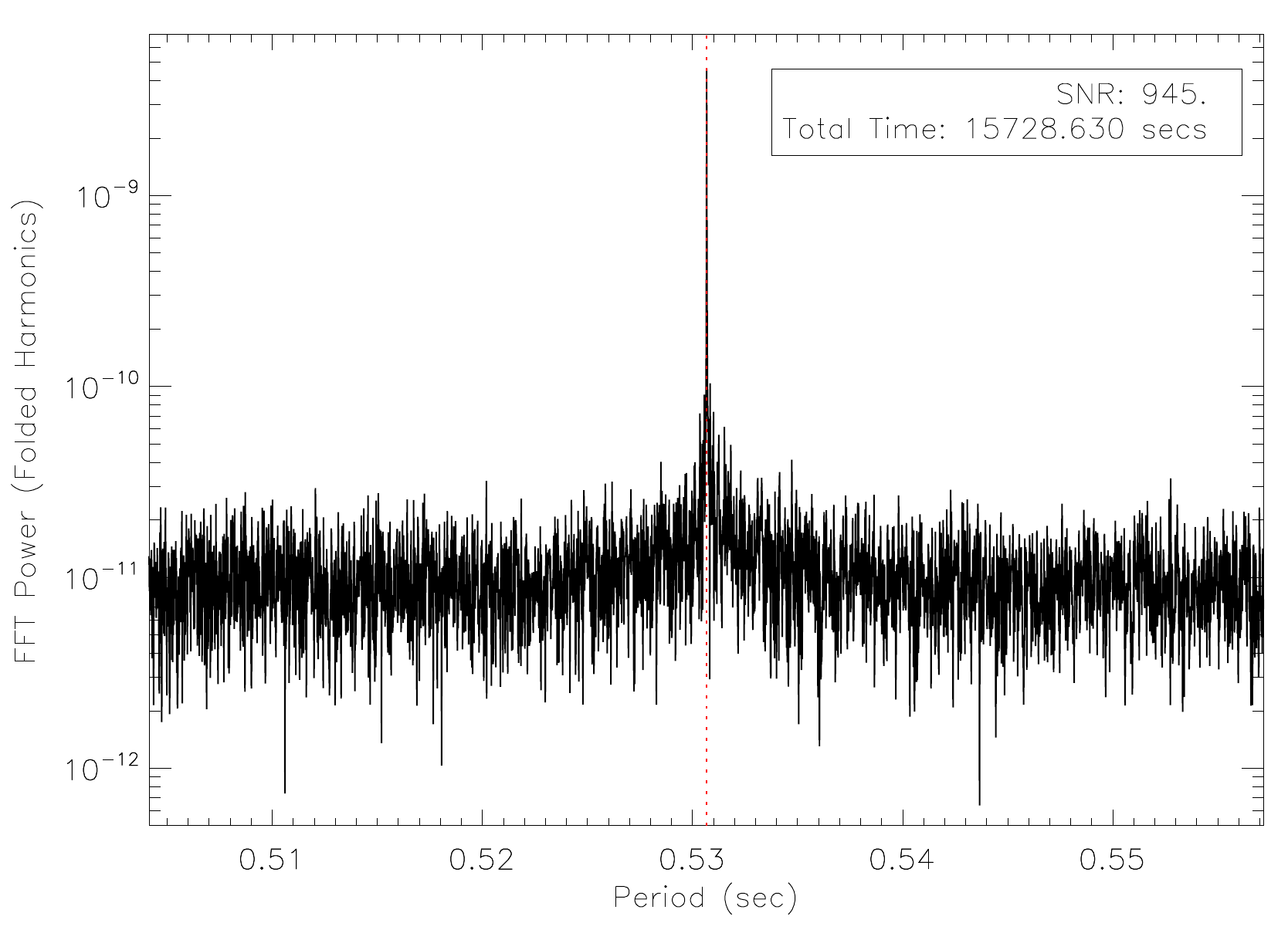}}\\
\vspace{-0.2em}
{\put(0,170){(C)}\includegraphics[width=0.5\textwidth]{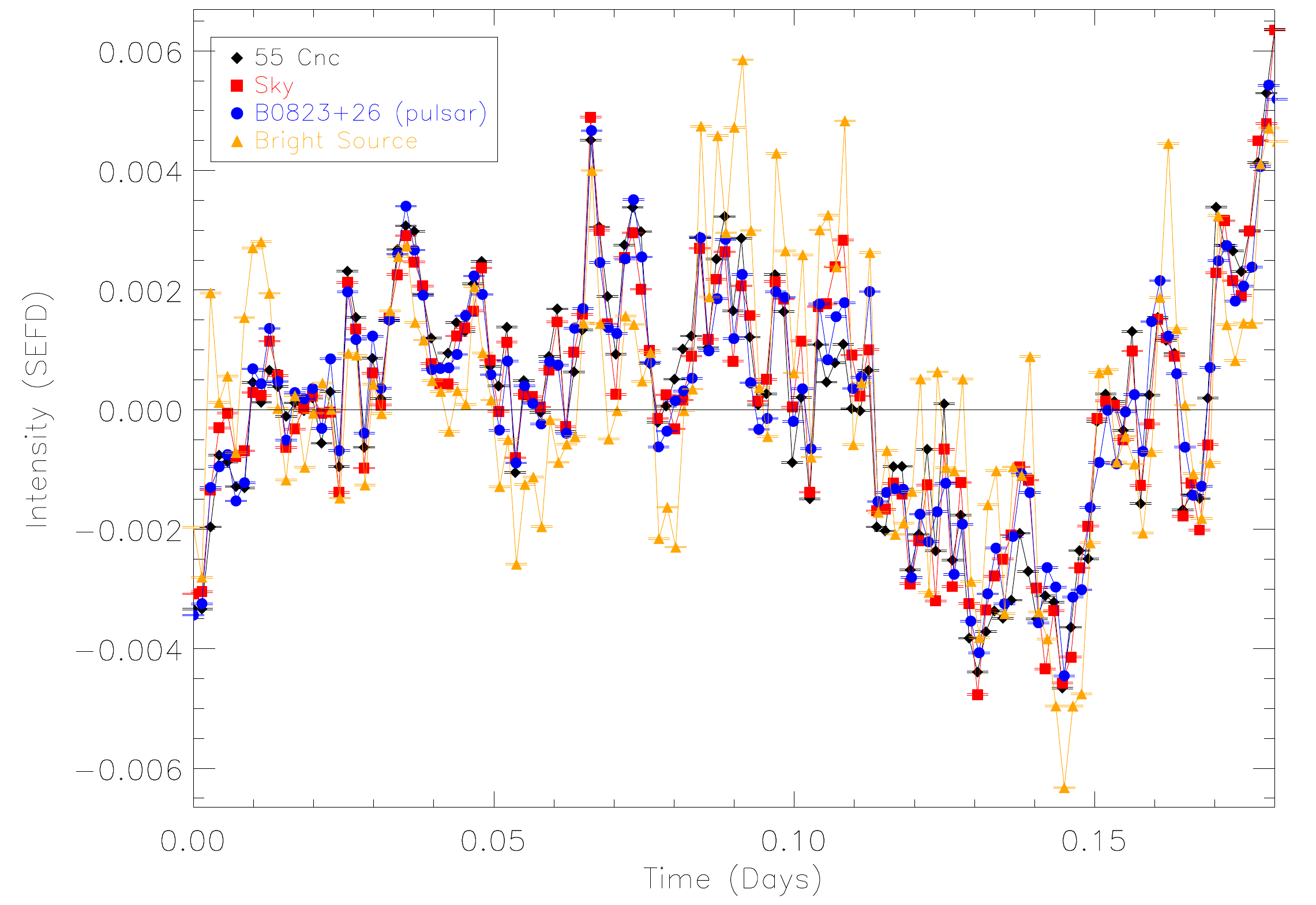}}&
{\put(0,170){(D)}\includegraphics[width=0.5\textwidth]{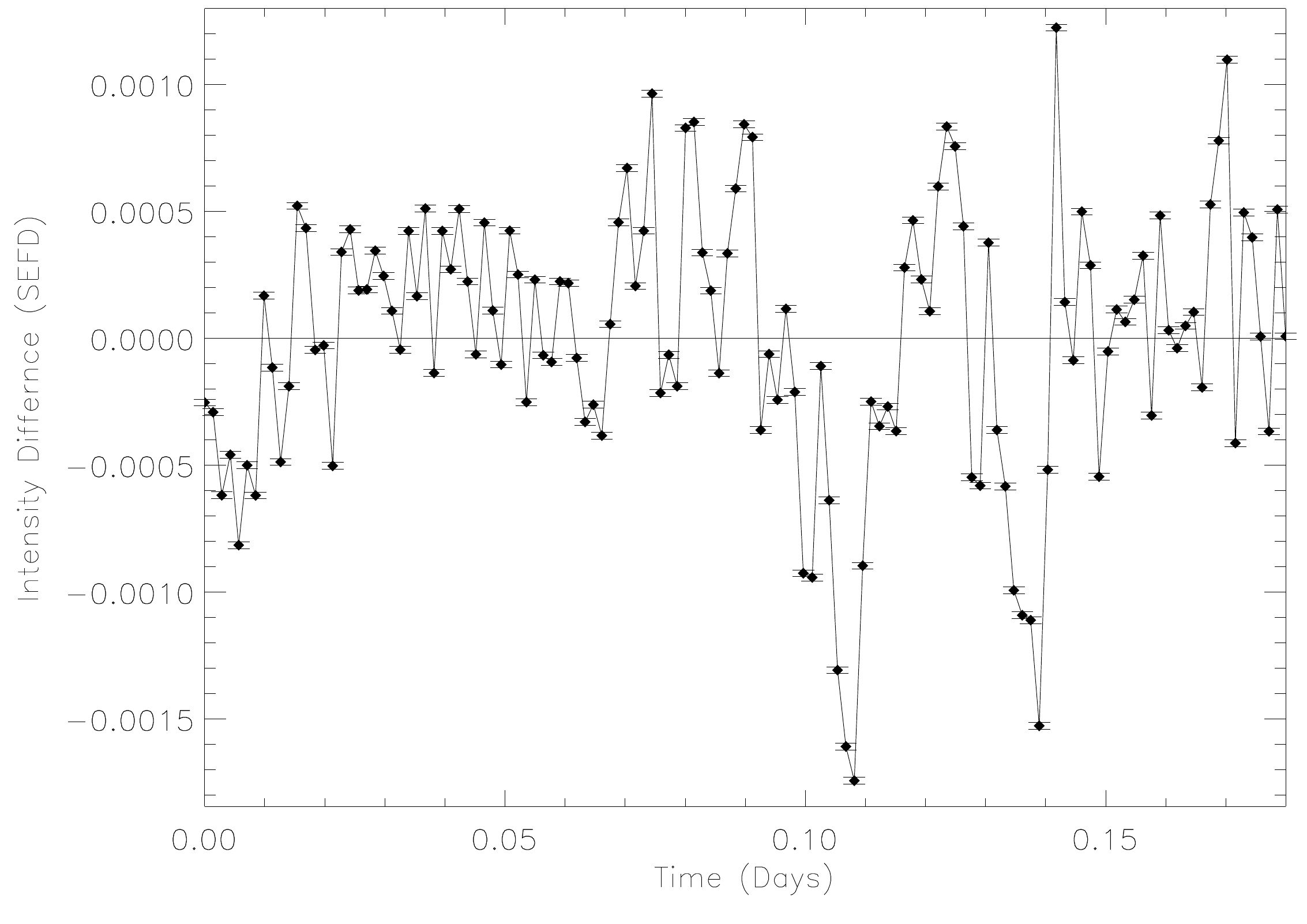}}
\end{tabular}
  \caption{
  Results of the pipeline for Obs~$\#$1. \textbf{(A)} An example of the normalized dynamic spectrum before and after RFI mitigation. The RFI has been masked and set to the median of the data not containing RFI. \textbf{(B)} FFT of the pulsar B0823+26. The known period of the pulsar is marked as a dashed red line and we recover the known period with a signal-to-noise ratio of $\sim$945. \textbf{(C)} Time-series of the intensity of all the beams. The intensity was found by integrating over the entire bandwidth ($47$ MHz) and rebinning to 2 minute intervals. The beams are 55 Cnc (black), sky (red), B0823+26 (blue), and bright source (orange). All curves have been subtracted by one. All beams are similar but we observe scintillation in the bright source. \textbf{(D)} Time-series of the intensity difference of the 55 Cnc and sky beam. These two beams are similar and oscillations due to the ionosphere are visible. No bursty radio emission from 55 Cnc is seen. 
  }
  \label{fig:results}
  \end{center}
\end{figure}

RFI mitigation is the most crucial step in the pipeline since RFI dominates the signal in the low-frequency data and hinders detection of faint astrophysical signals. The RFI mitigation pipeline consists of the following steps: (1) divide the raw data into slices of 42 seconds (4000 spectra), (2) find the frequency response function and divides the data by this function, (3) find RFI, and (4) save the location of the RFI into a mask, an array the same dimensions as the data with a value of 0 (polluted pixels) and 1 (clean pixels) weight. An example of the RFI mitigation can be found in panel A of Figure \ref{fig:results}.  It can be seen that the pipeline is very efficient at finding and masking the brightest RFI. Examining the dynamic spectrum, the integrated spectrum, the time-series (panel C of Figure \ref{fig:results}), and the fast Fourier transform (FFT) of the pulsar (Section \ref{sec:results}; panel B of Figure \ref{fig:results}) after RFI mitigation shows that minimal RFI is left-over in the data. In total, we mask out $\sim$3$\%$ of the data. The standard deviation of the data after RFI mitigation decreases by a factor of $\sim$100.

Step (1) and step (2) are required because the dynamic spectrum should not contain any large-scale variations in time and frequency in order to correctly apply step (3). Using only 42 second slices (4000 spectra) in step (1) guarantees that any changes in time are small. Step (2) allows for both the correct identification of RFI located on the outer edges of the response function and not introducing false-positive detections of RFI near the peak response of the telescope. The frequency response function in step (2) is created using the 10$\%$ quantile of the distribution of intensities at each frequency because the 10$\%$ quantile is relatively robust against RFI. 

Step (3) consists of 4 RFI mitigation techniques (Offringa et al. 2010, Offringa 2012, Zakharenko et al. 2013, Vasylieva 2015, and references therein) combined together for optimal efficiency and processing time: PATROL (Pulsars And TRansients Overall Lookup), ZURFIM (Ze Ultimate RFI Mitigator), SUM (SumThreshold), and Polluted pixel EXpansion (PEX). Each of these techniques uses sigma thresholding (above which the samples are flagged) in the time-frequency domain. After each RFI method is ran, the code updates the mask, and then only runs the next RFI method on the remaining good data. The 4 RFI techniques are described in greater detail below.  
The first two methods PATROL and ZURFIM both flag whole bad frequency channels and time intervals in the dynamic spectrum (Zakharenko et al. 2013, Vasylieva 2015). These techniques are useful for finding RFI caused by broadcasting radio stations either local or reflected from the ionosphere which results in contaminated  frequencies for the entire time interval. Additionally, they can also find short wide-band RFI spikes at all frequencies caused by telescope equipment, vehicles, lightning, or other natural sources. These programs identify RFI using values of the mean ($\mu$) and standard deviation ($\sigma$) of each channel. PATROL uses the spectrum integrated over all time to flag bad frequencies and the time-series integrated over all frequencies to flag bad channels, whereas ZURFIM uses the spectrum for each time channel and the time-series for each frequency channel and loops through all times and frequencies. Any frequency channel or time interval with a $\sigma$ above the sigma thresholding value is flagged (PATROL and ZURFIM do not necessarily have the same thresholding value). The process is iterative until there are no more peaks to exclude. 

The third method SUM (Offringa et al. 2010, it is also used as the default RFI mitigation pipeline for LOFAR visibilities) is designed to only flag small patches in the time or frequency direction. A combination of $n$ samples is entirely flagged if its average exceeds the threshold $T_{n}$ (in units of $\sigma$). $T_{n}$ is equal to $
    T_{n} = T_{1}/a^{\log_{2}n}$,
where $T_{1}$ is the threshold for a single pixels ($T_{1}$ = 10 for default), $a$ is an empirical coefficient ($a =1.5$), and $n$ is the size of the sliding window (in powers of 2). The program runs multiple times depending on how many values of $n$ are supplied. The time and frequency directions are run independently and then the masks are multiplied together at the end. 

The fourth RFI method PEX is very simple. The method just expands the polluted pixels in the mask by a certain number of pixels in both the time and frequency direction. The program also has the option to only expand patches of interconnected bad pixels of a certain size. This method is useful because strong RFI may contain weaker edges that might be missed by the other methods. PEX is very similar to the Scale Invariant Rank technique (SIR; Offringa 2012) but is simpler and thus computationally much faster. 

\subsection{Time-Frequency Telescope Response} \label{sec:tf}
The next part of the pipeline is designed to find the time-frequency response of the telescope. During the development of the pipeline, we found that the frequency response of the telescope changes with time and is different for every beam. These variations are caused by the shape of the beams changing while tracking the different sources. In the pipeline we (1) apply the RFI mask from Section \ref{sec:RFI_mask} to the raw data, (2) rebin each slice in the time dimension by a factor of 10, (3) find a second order polynomial fit at each frequency over the entire rebinned and RFI masked data, and (4) create and save the 2-d time-frequency response surface made from these polynomial fits. Step (2) is used to avoid biasing the polynomial fit with any short-term variability and to reduce computational memory. One of the limitations of the above procedure is that any constant or slowly varying signal will change the 2-d time-frequency response function, and thus get removed when we normalize by this function. However, this is not a problem for bursty signals such as the one we expect from an exoplanet.

\subsection{Apply Corrections}
The last part of the pipeline applies the corrections to the data. For every beam except the pulsar beam we (1) apply the RFI mask from Section \ref{sec:RFI_mask} to the raw data, (2) flatten the RFI masked data by the 2-d time-frequency response surface from Section \ref{sec:tf} (now the data are in units of the SEFD), (3) rebin the calibrated and RFI masked data in time and frequency, and (4) save the calibrated, RFI masked, and rebinned data. For the pulsar beam, we perform step (1) and (2) above, (3) de-disperse the data using the pulsar's known dispersion measure and rebin in frequency, and (4) save the calibrated, RFI masked, de-dispersed, and rebinned data. For all the beams excluding the pulsar beam we rebin to a spectral and time resolution of 45 kHz and 1 second, respectively. For the pulsar beam, the data is rebinned to a spectral resolution of 1 MHz and kept at a time resolution of 10 msec. 

\section{Results} \label{sec:results}
With the calibrated and RFI masked data we can now search for bursty astrophysical signals. The main signals in our data are the pulsar, scintillation of the bright source, ionospheric variations in the sky beam, and any emission from the exoplanet. Each of these signals is described below. 

The pulsar (B0823+26\footnote{All physical information for B0823+26 was taken from the ATNF Pulsar Catalogue (Manchester et al. 2005).}) beam is useful because the signal is faint, astrophysical in nature, and can be used to test the reliability of the pipeline. In order to detect B0823+26 we (1) perform an FFT on the calibrated, RFI masked, de-dispersed, and rebinned data and (2) add together the 6 first harmonics in the power spectrum. The FFT was computed using data from 35-49 MHz. If we perform an FFT on the de-dispersed data without masking the RFI, the pulsar is not detected. Thus RFI mitigation is a necessary step in the analysis. We use the pulsar beam to test different tunable parameters in the pipeline such as the thresholding values for the RFI mitigation. The pipeline RFI-mitigation parameters that maximized the FFT power were used for all the beams. An example of an FFT performed on Obs~$\#$1 can be found in panel B of Figure \ref{fig:results}. The pulsar is detected at its known period with a very high signal-to-noise ratio ($SNR_{FFT}$) of $\sim$945.

Next, we can search for astrophysical signals in our data by plotting the time series of the dynamic spectrum integrated over all frequencies. The time-series of all the beams for Obs~$\#$1 can be found in panel C of Figure \ref{fig:results}. Each beam's intensity was found by integrating over the entire bandwidth ($47$ MHz) and rebinning to 2 minute intervals. The intensity is in units of the SEFD and to emphasize the variations we also subtract 1 from the data. The SEFD at 30 MHz is approximately equal to the sky background but at 70 MHz it is $\sim$2x the sky. Scintillation of the bright source is seen in the time-series (orange curve in panel C of Figure \ref{fig:results}). We can only see fluctuations in the bright source's flux and not its average flux due to the way we constructed the time-frequency normalization curve (see Section \ref{sec:tf}). The target and sky beams behave roughly the same suggesting similar but not identical ionospheric conditions. Panel D of Figure \ref{fig:results} shows the time-series of the difference between the black (target beam) and red (sky beam) curves of panel C. The variations in the 55 Cnc and sky beam are due to changes in the ionosphere. There are no positives peaks above 2$\sigma$, therefore, we do not detect any emission from the exoplanet. However, there are negative peaks above 2$\sigma$. This result suggests that we need to have 2 OFF sky beams in future observations to verify that a detection is not a false-positive.

\section{Discussion} \label{sec:discussion}
Even though we do not detect any radio emission from the 55 Cnc system, we can still place upper limits on its radio emission. This can be done independently using both the FFT of B0823+26 (panel B of Figure \ref{fig:results}) and the time-series intensity difference of 55 Cnc to the sky beam (panel D of Figure \ref{fig:results}). 

The upper limit using the pulsar is obtained using the following procedure. The sensitivity of the observations ($\sigma\sim S_{pulsar}/SNR$) can be estimated using the SNR of the pulsar in the time domain and its intrinsic flux measured at the wavelengths observed ($S_{pulsar}$). However, the $SNR_{FFT}$ in the Fourier domain is not the same as the SNR in the time domain. Therefore, we need to determine a conversion factor between the two. We run simulated pulsar data (random Gaussian noise + pulse, same time and frequency resolution of our data, same time interval) and adjust the pulse/noise ratio to reproduce the $SNR_{FFT}$ of the FFT. We find that a pulse/noise ratio of 0.15 (pulse amplitude = 0.15$\sigma$) corresponds to the observed SNR$_{FFT}\sim$945 in the FFT (~$SNR_{FFT}~\sim~6330~\ SNR$~). Therefore, the sensitivity of the observations over any time scale ($\tau$) is 
\begin{equation}
\sigma(\tau) \approx  6330 \frac{S_{pulsar}}{SNR_{FFT}} \left(\frac{0.0105 \ \text{sec}}{\tau} \right)^{1/2}. \label{eq:sigma}
\end{equation}
In order to obtain the flux of B0823+26 ($S_{pulsar}$), we have taken a series of observations using the LOFAR station FR606 in stand-alone mode. The observations were taken in the LBA band (we used data from 50-90 MHz) and flux-calibrated using the method described in Kondratiev et al. (2016); however, we used the beam model 'Hamaker-Carozzi' instead of the 'Hamaker' beam model. With this, the median flux of B0823+26 over 13 observations was measured as 1210$\pm$150 mJy. Using equation (\ref{eq:sigma}), we find a 3-sigma upper limit of 230 mJy for 55 Cnc for an integration time of 2 minutes and over the entire bandwidth. The advantage of using the FFT for the noise calculation is that any effects (ionosphere, left over instrumental systematics, low-level RFI) not periodic with the pulsar's period are not taken into account. The 1-sigma sensitivity estimated for the pulsar ($\sim$76 mJy) is a factor of $\sim$5 higher than the theoretical sensitivity found using the LOFAR calibration data (16 mJy; Section \ref{sec:LOFAR}) and this factor likely arises from imperfect coherent addition of the station signals and the RFI flagging of data (LOFAR Astronomer's website on Beam Formed Mode\footnote{http://www.astron.nl/radio-observatory/observing-capabilities/depth-technical-information/major-observing-modes/beam-form}). 

\begin{figure}[htb!]
 \begin{center}
 \includegraphics[width=0.7\textwidth]{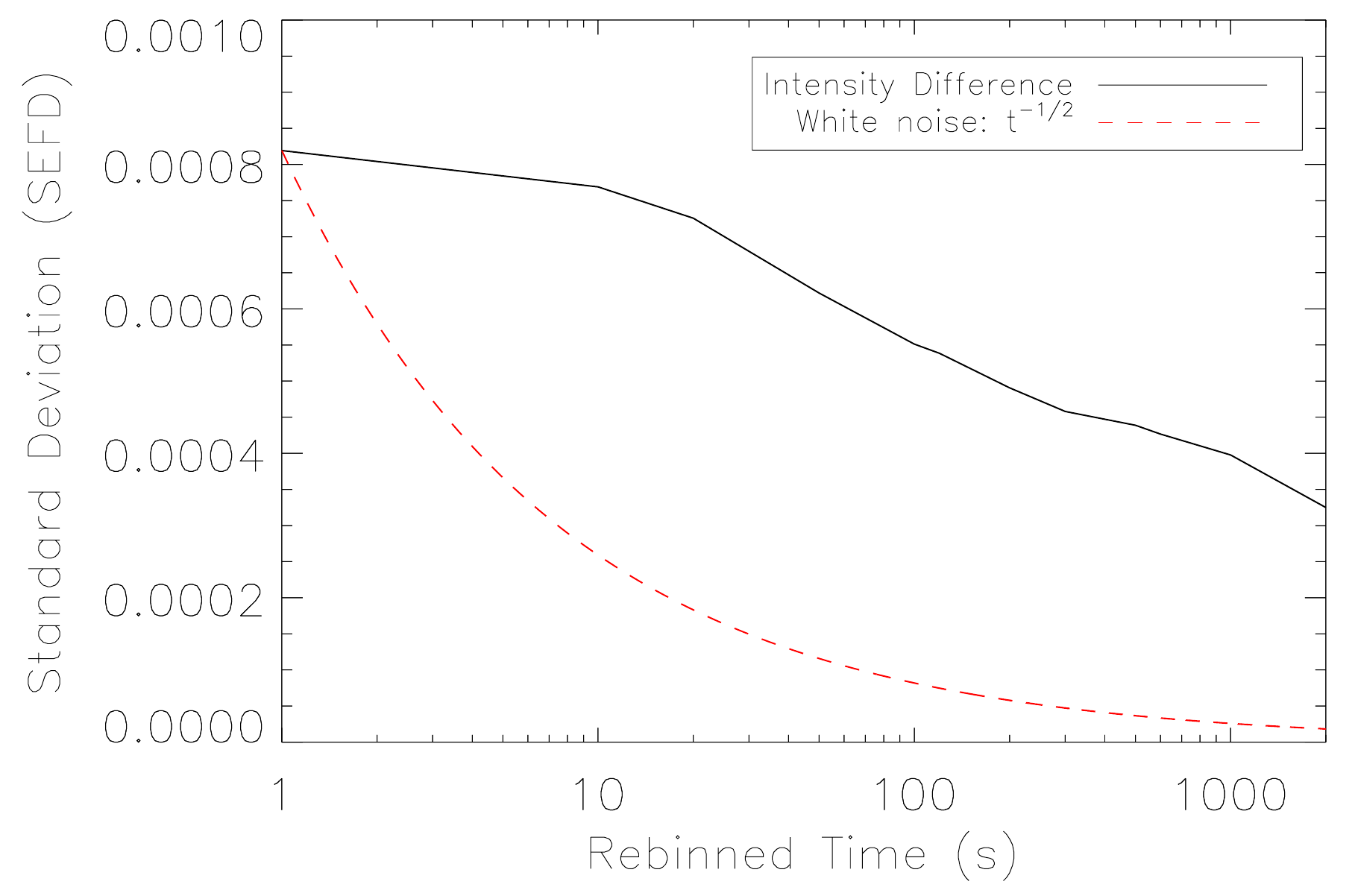}
  \caption{Standard deviation of the intensity difference between the target and sky beam for different rebin times (starting at 1 sec). The red dashed curve shows the theoretical white noise curve assuming the noise decreases by $t^{-1/2}$ (the curve starts at the measured standard deviation for a 1 sec rebin time).  
  }
   \label{fig:results2}
 \end{center}
 \end{figure}

Next, we can estimate the upper limit using the time-series of the intensity difference between the target and sky beam. The standard deviation in this time-series is $\sim$0.0005 of the SEFD (panel D of Figure \ref{fig:results}; Figure \ref{fig:results2}). The SEFD from 30-70 MHz for 24 LOFAR stations is 1.7 kJy (the SEFD from one station is 40 kJy; van~Haarlem et al. 2013). Therefore, the 1-sigma and 3-sigma sensitivity from these observations would be $\sim$850 mJy and 2.6 Jy, respectively. This 1-sigma limit is $\sim$11 times greater than the sensitivity limit derived using the pulsar ($\sim$76 mJy) and $\sim$50 times greater than the thermal noise ($\sim$16 mJy; Section \ref{sec:LOFAR}). From this factor of 11, a factor $\sim\sqrt{2}$ comes from the fact that Figure \ref{fig:results}D is a difference between the fluctuations of the target and sky beams. Part of the factor of $11/\sqrt{2}\sim8$ may be caused by observing close to the Galactic plane, that is brighter than the high Galactic latitudes. Although 55 Cnc is not in a very bright region of the galactic plane (see LFmap model from Polisensky 2007). We believe that a large part of this factor of $\sim$8 is due to the different fluctuations of the ionosphere between the two beams. This suggests that the ionosphere substantially varies at an angular scale of a few degrees. Rebinning to different times does improve the standard deviation but only slightly and not with a $t^{-1/2}$ white noise dependence (Figure \ref{fig:results2}). Using the time-averaging method (e.g. Pont et al. 2006, Turner et al. 2016), we find that there is a substantial amount of red noise (RMS of red noise $\sim$ 0.5 RMS of white noise) in the time-series. Therefore, this indicates that non-Gaussian ionospheric variations are present in the data over many timescales (at least between 1 and a 1000 seconds). The 3-sigma upper limit is also 25 times larger than the theoretically predicted flux density of $\sim$100 mJy for 55~Cnc~e  (Grie{\ss}meier et al. 2007, Nichols et al. 2016, Grie{\ss}meier 2017). Hence, with this upper limit it is not yet possible to put strong constraints on the theoretical emission models. 

Our results suggest that LOFAR LBA beam-formed observations may not be sensitive enough to detect exoplanetary radio emission due to many effects (large SEFD, non-coherent summation, differential ionospheric variations) with the current setup. The large ionospheric variations severely limit the detection capability using only 2 beams. Therefore, three beams (ON and 2 OFF) may be necessary to verify any possible detections against false-positives. Future exoplanet beam-formed observations with LOFAR will be performed with this new setup. Finally, we will more accurately quantify the sensitivity that LOFAR LBA beam-formed observations can reach in an upcoming study using LOFAR observations of Jupiter scaled such that it simulates exoplanetary radio emission (Turner et al. in prep). 
 
\section{Conclusion}
In this paper we present LOFAR Low Band Antenna beam-formed observations of the exoplanetary system 55 Cnc at 26-73 MHz. This is the first published paper on the search for exoplanet radio emissions using beam-formed observations from LOFAR. We created an automatic pipeline to flatten the LOFAR data by the time-frequency response of the telescope, find and mask RFI, and to search for astrophysical signals in our data. During the observations four beams were recorded simultaneously on 55 Cnc, a patch of nearby ``empty'' sky, the nearby pulsar B0823+26, and a bright constant radio source. The extra beams are used to monitor the time-frequency response of the telescope and ionospheric variability, and to verify the reliability of the pipeline. The pipeline was extensively tested and we found the data to be stable and sensitive enough to detect astrophysical signals from the pulsar and scintillation from the bright source. 

Initial analysis of 4 hours of LOFAR data do not show an exoplanet signal. We find a 3-sigma upper limit for the 55 Cnc system of 230 mJy using analysis of the pulsar to estimate the sensitivity and 2.6 Jy using the difference between the integrated time-series of the target and sky beam. These upper limits are a factor of $\sim$5 and $\sim$50 greater than the theoretical sensitivity. The factor of 5 for the pulsar is likely due to imperfect coherent addition of station signals (that also applies to the target-sky beam) and possibly residual RFI. The additional factor of $\sim$11 that affects the target-sky beam is attributed for a large part to large-scale differential variations of the ionosphere between the two beams. This result suggests that the ionosphere substantially varies at an angular scale of a few degrees. Therefore, in all future exoplanet beam-formed observations with LOFAR we will observe with three beams (one ON beam and two OFF beams) to decrease the detection of false-positives. Additionally, 55 Cnc is located on the Galatic plane which likely contributions an additional factor to the sensitivity calculation.

The findings in this study suggest that LOFAR LBA beam-formed observations may not be sensitive enough to detect exoplanetary radio emission or place strong constraints on model predictions. In the future, we will use our pipeline to analyze the full data set of the 55 Cnc observations, more accurately determine the sensitivity LOFAR can reach using Jupiter as a proxy for exoplanetary radio emission, and search for radio emission from other exoplanets predicted to have detectable radio emission. Finally, the techniques in this paper can be used to analyze beamformed data from future ground-based low-frequency radio telescopes (NenuFAR, LOFAR 2.0, SKA) or reconstructed dynamic spectra from the visibilities of imaging data (Loh et al. in prep).

\section*{Acknowledgements} 
J. Turner was funded by the National Science Foundation Graduate Research Fellowship under Grant No. DGE-1315231. We also thank PNP and PNST for their support. This research has made use of the Exoplanet Orbit Database, Exoplanet Data Explorer at exoplanets.org (Han et al. 2014), Extrasolar Planet Encyclopaedia (exoplanet.eu) maintained by J. Schneider (Schneider et al. 2011), NASA's Astrophysics Data System Bibliographic Services, and the ATNF Pulsar Catalogue (Manchester et al. 2005) located at http://www.atnf.csiro.au/research/pulsar/psrcat. We thank Louis Bondonneau for help with the flux calibration of B0823+26. Finally, we thank the referees, whose comments helped to improve this paper.

\section*{References}
\everypar={\hangindent=1truecm \hangafter=1}
Grie{\ss}meier,~J.--M., P.~Zarka and H.~Spreeuw, Predicting low-frequency radio fluxes of known extrasolar planets, \textsl{Astron. Astrophys.}, \textbf{475}, 359--368, 2007.

Grie{\ss}meier,~J.--M., Detection Methods and Relevance of Exoplanetary Magnetic Fields, in \textsl{Characterizing Stellar and Exoplanetary Environments}, edited by Lammer, H., and Khodachenko, M., Springer International Publishing, Cham, 213--237, 2015. 

Grie{\ss}meier ,~J.--M., Detection Methods and Relevance of Exoplanetary Magnetic Fields, in \textsl{Planetary Radio Emissions VIII (this volume)}, Austrian Academy of Sciences Press, Vienna, 2017.

Han E., Wang S. X., Wright J. T., Feng Y. K., Zhao M., Fakhouri O., Brown J. I., and Hancock C., Exoplanet Orbit Database. II. Updates to Exoplanets.org, \textsl{PASP}, \textbf{126}, 827, 2014.

Hess,~S.\,L.\,G., and P.~Zarka, Modeling the radio signature of the orbital parameters, rotation, and magnetic field of exoplanets, \textsl{Astron. Astrophys.}, \textbf{531}, A29, doi:10.1051/0004-6361/201116510, 2011.

 Kondratiev, V.~I., et al. (31 co--authors), A LOFAR census of millisecond pulsars, \textsl{Astron. Astrophys.}, \textbf{585}, A128, 2016.


Lecavelier~des~Etangs,~A., S.\,K.~Sirothia, Gopal--Krishna, and P. Zarka, Hint of 150~MHz radio emission from the Neptune--mass extrasolar transiting planet HAT--P--11b, \textsl{Astron. Astrophys.}, \textbf{552}, id.A65, 6 pp., 2013.

Manchester, R. N., Hobbs, G. B., Teoh, A., and Hobbs, M., The Australia Telescope National Facility Pulsar Catalogue, \textsl{Astrophys. J.}, \textbf{129}, 1993--2006, 2005.

Nichols, J.~D., and Milan, S.~E., Stellar wind-magnetosphere interaction at exoplanets: computations of auroral radio powers, \textsl{MNRAS}, \textbf{461}, 2353--2366, 2016.

Offringa., A. R., Algorithms for Radio Interference Detection
and Removal, PhD thesis, University of Groningen, Groningen, Netherlands, 2012.

Offringa, A.~R., {de Bruyn}, A.~G., {Biehl}, M., {Zaroubi}, S., {Bernardi}, G., and {Pandey}, V.~N., Post-correlation radio frequency interference classification methods, \textsl{MNRAS}, \textbf{405}, 155, 2010.

Polisensky, E., LFmap: A Low Frequency Sky Map Generating Program, \textsl{Long Wavelength Array (LWA) Memo Series}, \textbf{111}, 2007.

Pont F., Zucker S., and Queloz D., The effect of red noise on planetary transit detection, \textsl{MNRAS}, \textbf{373}, 231, 2006.

Schneider,~J., Dedieu, C., Le Sidaner, P., Savalle, R., and Zolotukhin, I., Defining and cataloging exoplanets: the exoplanet.eu database, \textsl{Astron. Astrophys.}, \textbf{532}, A79, 2011.

Sirothia, S.~K., {Lecavelier des Etangs}, A., {Gopal-Krishna}, {Kantharia}, N.~G., and {Ishwar-Chandra}, C.~H., Search for 150 MHz radio emission from extrasolar planets in the TIFR GMRT Sky Survey, \textsl{Astron. Astrophys.}, \textbf{562}, A108, 2014.

Turner, J.D, et al., (42 co--authors), Ground-based near-UV observations of 15 transiting exoplanets: constraints on their atmospheres and no evidence for asymmetrical transits, \textsl{MNRAS}, \textbf{459}, 789, 2016.

van~Haarlem,~M., et al. (200 co--authors), LOFAR: The LOw-Frequency ARray, \textsl{Astron. Astrophys.}, \textbf{556}, id.A2, 53 pp., 2013.

Vasylieva, I., Pulsars and transients survey, and exoplanet search at low-frequencies with the UTR-2 radio telescope: methods and firrst results, PhD thesis, Paris Observatory, Paris, France, 2015. 

Young, N.~J., Stappers, B.~W., Weltevrede, P., Lyne, A.~G., and Kramer, M, On the pulse intensity modulation of PSR B0823+26, \textsl{MNRAS}, \textbf{427}, 114, 2012.

Zakharenko,~V.\,V., I.\,Y.~Vasylieva, A.\,A.~Konovalenko, O.\,M.~Ulyanov, M.~Serylak, P.~Zarka, J.--M.~Grie{\ss}meier, I.~Cognard, and V.\,S.~Nikolaenko, Detection of decametre--wavelength pulsed radio emission of 40 known pulsars, \textsl{MNRAS}, \textbf{431}, 3624--3641, 2013.

Zarka, P., Lazio, J., and Hallinan, G., Magnetospheric Radio Emissions from Exoplanets with the SKA, in \textsl{Advancing Astrophysics with the Square Kilometre Array (AASKA14)}, 120, 2015.







\end{document}